\documentclass[12pt]{article}
\usepackage[dvips]{graphicx}
\usepackage{pdproc}

  \makeatletter 
  \def\@cite#1{[#1]} 
  \makeatother    
\newcommand{\beq}{\begin{equation}}
\newcommand{\eeq}{\end{equation}}
\newcommand{\beqs}{\begin{eqnarray}}
\newcommand{\eeqs}{\end{eqnarray}}

\def\Tr{\mbox{Tr}\,}

\def\hbar{\hspace{0pt}\raisebox{1pt}{$-$} \hspace{-7pt} h}

\def\5{\overline 5}

\begin{document}

\renewcommand{\thefootnote}{\alph{footnote}}

\title{
 Composite Vector Mesons and the Little Higgs Mechanism
}

\author{ MAURIZIO PIAI}

\address{ 
Physics Department,
Sloane Laboratory,
Yale University\\
New Haven CT 06520, U.S.
\\ {\rm E-mail: maurizio.piai@yale.edu}}

\abstract{
I review a technique to embed vector mesons in the 
chiral Lagrangian of QCD, and  apply
it to more general  coset spaces, relevant for 
Little Higgs models. 
The implementation of heavy spin-1 fields
in Little Higgs models allows for a better control over
previously non calculable, ultra-violate sensitive quantities, such as the
Higgs couplings.  A relevant application is the study of vacuum alignment
in the $SU(6)/Sp(6)$ models.
}

\normalsize\baselineskip=15pt
\vskip0.8cm

In Little Higgs (LH)  models, the Higgs fields
are pseudo-Nambu-Goldstone bosons
(PNGS) of an approximate global symmetry. 
This is an effective field theory, providing a  description  valid up
to the cut-off scale $\Lambda\sim4\pi f$, where $f$ is the
symmetry-breaking scale.
The symmetry structure of the models is such that
the Higgs mass can be radiatively generated  
only by loop diagrams involving more than one of the symmetry-breaking couplings,
thus suppressing it in respect to its natural size.

Some relevant quantities are quadratically sensitive to the cut-off of the theory.
In~\cite{vectorLH}, it has been shown that calculability 
can be improved by including in the spectrum of the
effective theory heavy spin-1 fields, to be thought of as heavy resonances
of the underlying (strong dynamics) UV completion.
The mechanism proposed is based on the idea of
{\it hidden symmetry}~\cite{hidden}, and its use in 
the QCD chiral Lagrangian. A crucial role
is played by the {\it vector limit}~\cite{georgi},
as a point of enhanced symmetry of the models, and
by {\it locality} in the
theory-space language~\cite{deconstruction,son}.

\section{Vector Mesons and the QCD Chiral Lagrangian. }

At low energies, 2-flavor QCD is well described by an effective 
Lagrangian containing the PNGB's of the broken phase 
(identified with the physical pions), describing the fluctuations
along the broken generators in 
the coset $SU(2)_L\times SU(2)_R/SU(2)_V$.
An effective Lagrangian  contains in general 
 also explicit symmetry breaking terms, such as
gauge interactions ($U(1)_{\rm em}\subset SU(2)_V$),
mass terms (quark masses), and possible
Wess-Zumino terms~\cite{WZW}.
I focus here on gauge interactions, neglecting other 
symmetry breaking terms.
The symmetry structure of the Lagrangian can be 
represented by a 2-site and 1-link diagram in the
theory-space language, in which the $D=4$ Lagrangian
describes a $D=5$ theory, with a discrete fifth dimension. 
Locality 
means allowing only  nearest-neighbor
interactions in the fifth dimension.

To introduce in the spectrum the $\rho$ mesons of QCD, one
rewrites the theory as a 3-site 2-link model, with an additional 
gauged $SU(2)$, so that the coset-space structure
becomes $SU(2)_L\times SU(2)_{\rho} \times SU(2)_R / SU(2)_V$.
The Lagrangian contains
\begin{eqnarray}
\label{Eq: effLrhopi}
{\cal L}\,=\,\frac{f^2}{4}\,\left( \Tr|D^\mu\Sigma_L|^2
+\Tr|D^\mu \Sigma_R|^2
+\,\frac{\kappa}{2}\,\Tr|D^\mu(\Sigma_L\Sigma_R)|^2\right)\, +\cdots ,
\end{eqnarray}
plus the Yang-Mills terms
for the $U(1)$ and $SU(2)_\rho$ gauge fields. 
The  gauge bosons acquire mass via spontaneous symmetry breaking:
half of the $\pi_i$ fields contained in $\Sigma_{L,R}\equiv \exp(i\pi_{L,R}/f)$ 
become, in
unitary gauge, the longitudinal components of the massive $\rho$ fields.
The spectrum of the model contains then the physical pions, with
a decay constant $f_\pi = \sqrt{\frac{1+\kappa}{2}} f$, 
the massless photon with coupling $e$, and three $\rho$ mesons with masses
proportional to the gauge coupling $g_{\rho}$. 
The electromagnetic  $U(1)_{\rm em}$ is generated by a linear combination of 
the generators of the gauged $U(1)\subset SU(2)_L\times SU(2)_R$
and $SU(2)_{\rho}$, with  mixing angle
$\cos^2\theta \approx 
\Big(1  - \frac{e^2}{g_\rho^2} \Big)$.

One can use this Lagrangian to compute the Coleman-Weinberg (CW) 
1-loop potential~\cite{CW},
from which one can read the
splitting between neutral- and charged-pion masses 
in terms of the cut-off scale  $\Lambda$ and of the mass of $\rho$'s~\cite{cwpions}:
\begin{eqnarray}
\delta m^2_{\pi^\pm} \simeq \frac{3e^2}{16\pi^2 (1 + \kappa)}\Big(
\frac{\kappa\Lambda^2}{\cos^2 \theta}  +  m^2_{\rho^0} 
\log\frac{\Lambda^2}{m^2_{\rho^0} } + \cdots \Big).
\label{pionsplit}
\end{eqnarray}
The procedure can be generalized, including more sites 
and hence more spin-1 resonances.

Consider the limit $\kappa\rightarrow 0$.
The Lagrangian becomes local in the theory space language.
As a consequence (see Eq.~(\ref{pionsplit})),
the quadratic divergence disappears,
in favor of a milder logarithmic divergence.
In the local limit, the CW potential
is less UV-sensitive,
and observable quantities can be better understood in terms
of low-energy degrees of freedom.
Further, $\kappa \rightarrow 0$ represents a point of enhanced symmetry: 
the chiral symmetry is restored to a global $SU(2)^2$  connecting 
the physical pions and the longitudinal components of the $\rho$'s.
This is the vector limit introduced first in~\cite{georgi}. 
Ordinary QCD is probably not too close to the vector limit 
(one can estimate $\kappa \sim 1/3$).
In a  general effective theory,
locality and vector limit are not controlled by the same 
parameter, and have to be considered
as distinct properties.
 
\section{Vacuum Alignment in $SU(6)/Sp(6)$}

In LH models based on the
general coset $G/H$, 
heavy spin-1 states can be implemented
in analogy to the QCD case. 
This allows to discuss 
physically relevant quantities, such as the mass and couplings of the Higgs fields,
the scale of unitarity violation~\cite{unitarity} and vacuum alignment. I focus
here on the last issue, in the interesting example
of  $SU(6)/Sp(6)$~\cite{su6}. 

The $SU(6)$ global symmetry is broken by an
antisymmetric condensate to its $Sp(6)$ subgroup.
Two $SU(2)_{1,2}$ groups are gauged 
(I  neglect the abelian $U(1)$'s here), 
and
the spontaneous breaking leaves only the diagonal $SU(2)_L$ unbroken.
Of the PNGB's, three are eaten in the generation of the  mass of 
the $W^{\prime}$ gauge bosons, the others 
arrange in two Higgs doublets, one complex  singlet $\eta$
and possible axion fields.
Several variations of the original model 
have been discussed, in connection with precision electroweak physics~\cite{precisionsu6},
or in order to describe flavor physics~\cite{flavons}.

A simple analysis suggests that this choice of the vacuum is unstable:
vacuum alignment seems to favor the unbroken $Sp(6)$ to 
contain the two $SU(2)$'s~\cite{stability}. This can be seen from the fact that
the complex singlet field acquires a negative, 
quadratically divergent mass term in the 1-loop CW potential. 
But the quadratic divergence
implies that this is a UV-dominated, not calculable quantity.

\begin{figure}[h]
\begin{center}
\includegraphics*[width=10cm]{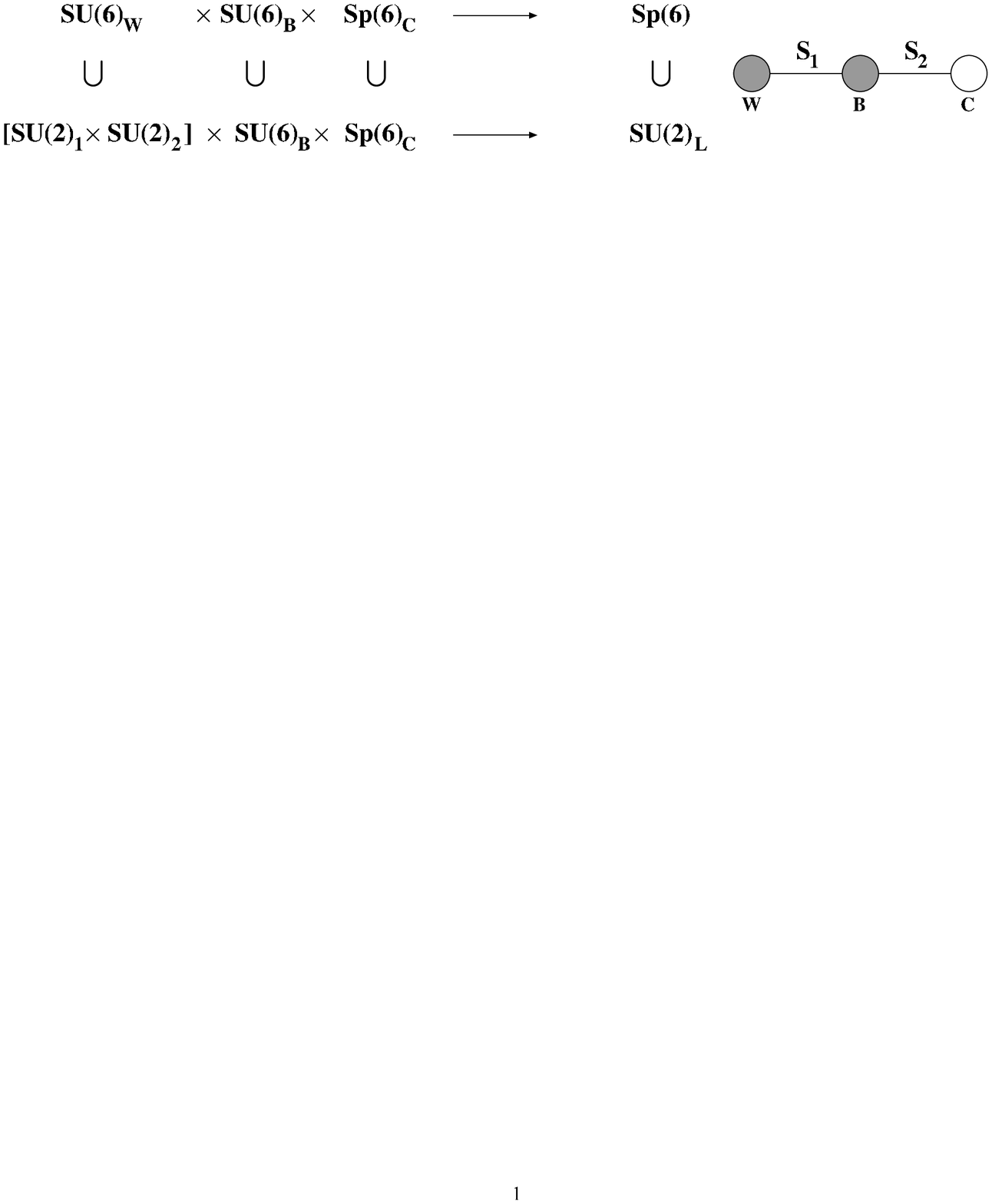}
\caption{Symmetry structure and diagrammatic representation of
the $SU(6)/Sp(6)$ model with inclusion of $\rho$, $a_1$ and $\rho^{\prime}$ states.
See text for details.
}
\label{fig:su6}
\end{center}
\end{figure}

Consider instead the case in which one introduces  in the spectrum the
first three spin-1 modes, namely two copies $\rho$ and $\rho^{\prime}$
of  techni-rho's and one copy of techni-$a_1$'s, with
$m_{\rho} < m_{a_1} < m_{\rho^{\prime}}$.
The construction is described in Fig.~\ref{fig:su6}.
The global symmetry is enlarged to 
a 3-site 2-link model,
in which $SU(6)_B\times Sp(6)_C$ is gauged, together with the $SU(2)^2$
subgroup of $SU(6)_W$. The breaking to $Sp(6)$ leaves
only an $SU(2)_L$ combination  unbroken.
The lowest-order terms in the Lagrangian (Yang-Mills kinetic terms are understood)  are
\begin{eqnarray}
\nonumber
{\cal L} &=&  
 f^2\Big( \Tr |D S_1|^2 + c^2 \Tr |D S_2|^2 \\
&+& \kappa_1 \Tr| D S_1 S_2|^2 + \frac{\kappa_2}{4} \Tr| D S_2 \Sigma_0 S_2^T|^2 
+ \frac{\kappa'}{4} \Tr| D S_1 S_2 \Sigma_0 S_2^TS_1^T|^2 
\Big)\label{su6vec}\,,
\end{eqnarray}
where $S_j=e^{i\,\pi_j/f}$.  The dimensionless constants $c^2$,
$\kappa_1$, $\kappa_2$, $\kappa^{\prime}$, as well as the 
gauge couplings $g_1$, $g_2$, $g_B$, $g_C$ in the covariant derivatives,
are determined by the (unknown) underlying dynamics.
In unitary gauge, all the components of $\pi_i$ in $Sp(6)$ disappear
to form longitudinal components of the $\rho$ and $\rho^{\prime}$ fields.
One linear combination of the PNGB's in the coset $SU(6)/Sp(6)$ is 
eaten to give mass to the $a_1$ fields. Of the other linear combination, three 
components give mass to the heavy $W^{\prime}$ bosons,
while the others are the light pseudo-scalar states to be identified with the 
scalar sector of the theory.

\begin{figure}[htb]
\begin{center}
\includegraphics*[width=8cm]{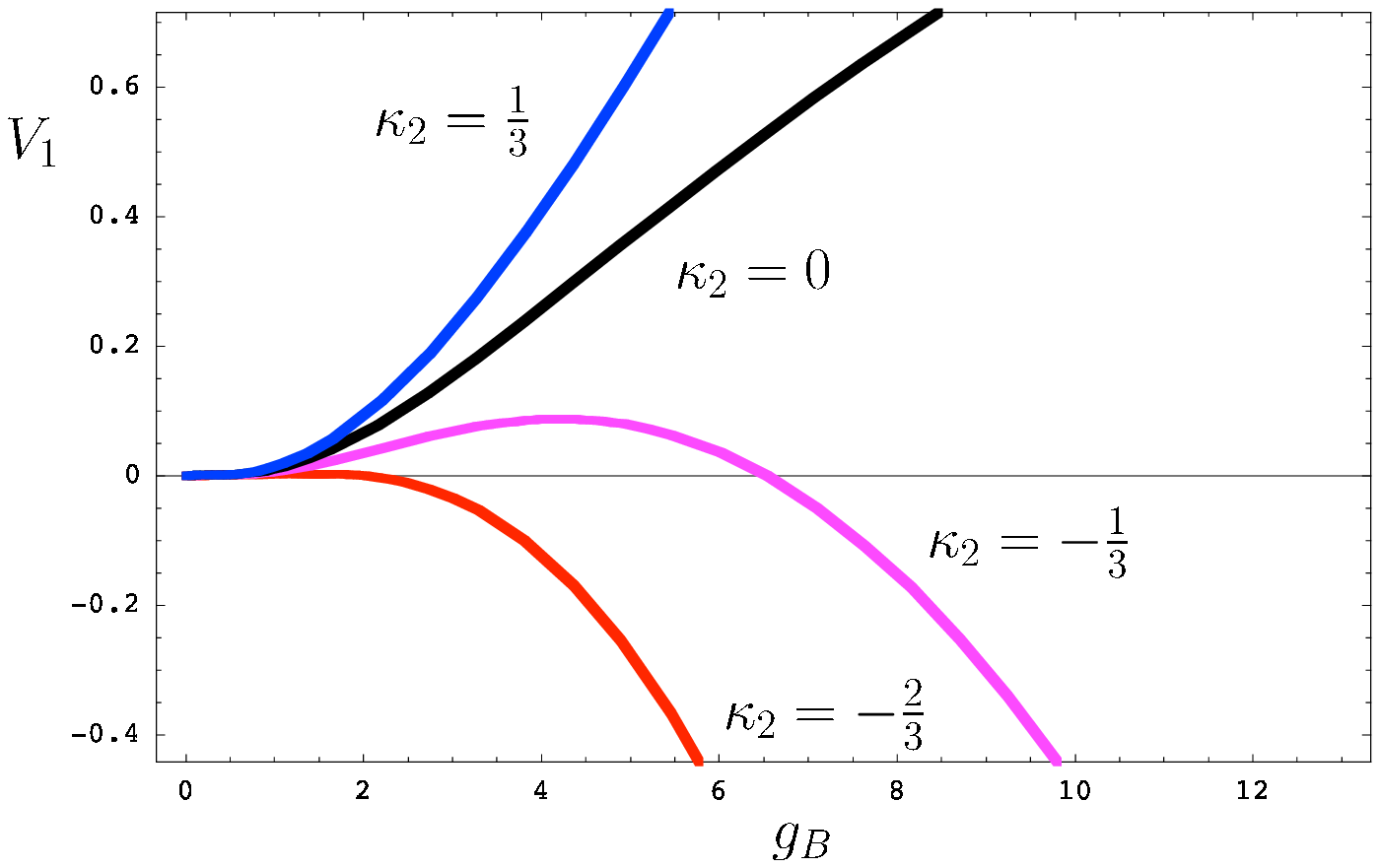}
\caption{The coefficient $V_1$ in the CW 1-loop potential
for the field $\eta$ in Eq.~(\ref{pot}) as a function of $g_B$
for different values of $\kappa_2$. All curves are computed for
$\kappa^{\prime}=\kappa_1=0$, $c^2=1$,
$g_1=0.65$, $g_2=3.3$, $g_C=5.2$.
}
\label{fig:eta}
\end{center}
\end{figure}

For simplicity, one can set $c^2=1$. 
The couplings $\kappa_1$ and $\kappa^{\prime}$ are non-local in the 
theory space description. 
Setting them to zero gives a finite 1-loop CW
potential for the $\eta$ field, which reads:
\beq
V(\eta)\,=\,V_1\,f^4\,\cos\left(\frac{|\eta|}{F_{\eta}}\right)\,,
\label{pot}
\eeq
where the  decay constant $F_{\eta}$ 
depends on $\kappa_2$ and $f$, and where constant terms and
higher order terms in the Fourier expansion have been neglected.
$V_1$ can be computed exactly.
The vector limit requires also $\kappa_2=0$.
How far from the vector limit must  the theory be, in order for the
(finite) coefficient of the  mass term for $\eta$ to turn positive, and the vacuum to be stable?
The answer is depicted in Fig.~\ref{fig:eta}.
The $\rho$ and $\rho^{\prime}$ contributions to $V_1$ are positive,
while $a_1$ contributes negatively.
For appropriate choices of the parameters, 
mixing effects become important, and the $\rho$'s contribute 
less than the heavier $a_1$'s, so that the vacuum may be stable.
To be more specific, for vanishing or positive values of $\kappa_2$, 
$V_1$ is a positive definite quantity and the vacuum unstable.
For negative values of $\kappa_2$, and for large values of $g_B$ 
(a region of parameter space where
 the theory is close to the strong coupling regime, and hence
 perturbative results have to be interpreted cautiously), $V_1$ turns negative,
thus showing on a computable example that
the naif expectation of vacuum alignment
can, in principle, be modified without invoking  the underlying dynamics.

\section{Acknowledgements}

I  thank the organizers for partial support,
and A.~Pierce and J.~Wacker for the collaboration
this work is based upon. 
Research supported 
by the grant DE-FG02-92ER-4074.

\bibliographystyle{plain}

\end{document}